\documentclass[twocolumn,showpacs,showkeys,superscriptaddress]{revtex4}

\usepackage{amsmath}
\usepackage{bbold}
\usepackage{amsfonts}
\usepackage{amssymb}
\usepackage{pbsi}
\usepackage[T1]{fontenc}
\usepackage{hyperref}
\usepackage{xcolor}
\usepackage{graphicx}
\usepackage{subfig}
\usepackage{float}

\usepackage{color}

\begin{document}

\title{Upshifted frequency of electromagnetic plasma waves due to reflecting gravitational waves acting as almost-luminal mirrors}

\author{Felipe A. Asenjo}
\email{felipe.asenjo@uai.cl (corresponding author)}
\affiliation{Facultad de Ingenier\'ia y Ciencias,
Universidad Adolfo Ib\'a\~nez, Santiago 7491169, Chile}

\author{Swadesh M. Mahajan}
\email{mahajan@mail.utexas.edu}
\affiliation{Institute for Fusion Studies, The University of Texas at Austin, Texas 78712, USA}

\begin{abstract}
We show that dispersive gravitational waves,  as a background spacetime, can reflect electromagnetic  waves in a plasma. This reflection upshifts the frequency of the reflected wave, being larger for low-frequency incident waves. This effect takes place when the gravitational wave background propagates almost at the speed of light, allowing it to behave similar to a luminal mirror to electromagnetic plasma waves. 
\end{abstract}

\maketitle

Frequency shift in propagating electromagnetic (EM) waves can be induced by a variety of processes - the conventional Doppler shift being the most common when there is a relative velocity between the emitter and the observer. Perhaps the most spectacular manifestation is the cosmological redshift originating in the expansion of the universe; it gives us the most trusted path to measure cosmic distances. No matter where frequency shift is observed, it stems from the interaction of the emitter, the observer and the properties of the medium through which the EM wave propagates. 

Such a phenomenon could also occur, for instance, when light propagates in a moving medium that can reflect it (the medium acts like a moving mirror); the frequency shift will depend on the velocity of the moving medium and of course whether it is approaching or receding from a wave front, i.e, if the "motion" of the medium is along or opposite to the direction of propagation of the EM wave. Of course, this effect must also takes place when the mirror velocity approaches to the speed of light. This 
was demonstrated a recent paper (see Ref.~\cite{Esirkepov}) where it was shown that he reflected  EM wave can experience an  upshift on its frequency when it propagates in a mirror medium that travels near (or equal) to the speed of light.

In this work we take a leap forward and generalize the concept of a mirror medium to include the spacetime filled with a gravitational wave propagating in a given direction.  The interaction of the electromagnetic plasma waves with such spacetime background can produce reflected electromagnetic plasma waves with an upshifted frequency that will depend on the gravitational wave profile. This will  occur when the gravitational wave is dispersive, but traveling at almost the speed of light.

This   mechanism for  frequency shift, due to a general wave-wave interaction,   could have obvious astrophysical and cosmological implications that we will discuss at the end of the paper.

We construct the simplest model to exhibit this class of wave-wave interaction. We study the electromagnetic wave dynamics in a medium  filled with a specified dispersive gravitational wave as the background for the plasma dynamics. As mentioned earlier, we expect the  gravitational wave to act as an effective almost-luminal mirror for electromagnetic plasma waves.

We deal with an ideal general relativistic one-component plasma fluid immersed in a gravitational field; the dynamic charged fluid component (electrons) move in a neutralizing background (provided by ions, for example). This electron fluid, with charge $q$, mass $m$ and density $n$,  can be described in the framework of a unified formalism \cite{mahajan, mahajan 2, asenjo2}, where electromagnetic and fluid dynamics are treated on the same footing. The unified formalism compresses the covariant plasma fluid dynamics into a single equation   \cite{mahajan, asenjo2}
\begin{equation}\label{plasanifs}
q U_\nu M^{\mu\nu}=T \nabla^\mu\sigma\, .
\end{equation}
where $\nabla_\mu$ is a covariant derivative for a general metric $g_{\mu\nu}$ with signature $(-,+,+,+)$, and with $\mu,\nu=0,1,2,3$. Here, 
 $U^\mu$ is the plasma fluid four-velocity, $T$ is the plasma temperature, $\sigma$ is its entropy, and  $M^{\mu\nu}=F^{\mu\nu}+({m}/{q}) S^{\mu\nu}$
is the  magnetofluid tensor  that unifies the electromagnetic field  $F^{\mu\nu}=\nabla^\mu A^\nu-\nabla^\nu A^\mu$ ($A^\mu$ is the electromagnetic four-potential), and $S^{\mu\nu}=\nabla^\mu \left(f U^\nu\right)-\nabla^\nu \left( f U^\mu\right)$, represents the thermal-vortical content of the plasma. The factor $f=h/mn$ , where $h$ is the plasma enthalpy density, contains the thermal-inertial effects, For a relativistic Maxwellian distribution, it is found that $f=K_3[(m)/(k_B T)]/K_2[(m)/(k_B T)]$, where $K_2$ and $K_3$ are the modified Bessel functions of the second
kind  \cite{mahajan, asenjo}. Here, $k_B$ is the Boltzmann constant and the speed of light $c$ is chosen to be unity. Eq. \eqref{plasanifs} is constrained by the continuity equation $\nabla_\mu(nU^\mu)=0$, and describes an isentropic plasma   $U_\nu \nabla^\nu\sigma=0$. Finally, to close the system, we need   Maxwell equations
\begin{equation}\label{maxwell}
\nabla_\nu F^{\mu\nu}=q n U^\mu\, .
\end{equation}

Interestingly, from the above model, we can find a simple exact propagation mode in curved spacetimes. 
For homentropic plasmas $\nabla^\mu\sigma=0$,  the simplest solution of Eq.~\eqref{plasanifs} is $M^{\mu\nu}=0$, which is readily  achieved when the generalized  vector potential $A^\mu+({m f}/{q}) U^\mu=0$; the latter equality determines the current that must go into Eq.~\eqref{maxwell}, yielding 
\begin{equation}
\nabla_\nu F^{\mu\nu}+\Omega_p^2 A^\mu=0\, ,
\label{plasmawaveequation}
\end{equation}
which is the equation that we must solve for electromagnetic waves in a medium immersed in curved spacetime. 
The constant  $\Omega_p=\omega_p/\sqrt{f}$ is the thermally corrected plasma frequency ($\omega_p=\sqrt{n q^2/m}$ at non-relativistic temperatures). 

The purpose of this work is to show that when a (near luminal) gravitational wave acts as a background for the electromagnetic plasma wave dynamics, the gravitational wave behaves as a mirror, allowing a total reflection of the electromagnetic wave, and  upshifting its frequency. In order to show this, let us start considering a dispersive (subluminal) gravitational wave as a background. A dispersive feature of a gravitational wave occurs when it  propagates in a massive medium \cite{forsberg,brodin,mendonca,bamba,asseo,chesters,gayer,polnarev,ingram,servin,mendonca2,silker}.
 The spacetime  metric is  $g_{\mu\nu}=\eta_{\mu\nu}+h_{\mu\nu}$, where $\eta_{\mu\nu}=(-1,1,1,1)$ is the flat spacetime metric, and  $h_{\mu\nu}$ is  the  gravitational wave  perturbation ($h_{\mu\nu}\ll \eta_{\mu\nu}$).  
For simplicity, we  consider a gravitational wave with only two nonzero components, propagating in the positive  $z$-direction, $h_{11}=-h_{22}=h(\chi)$ (with $h\ll 1$); the fields are functions only of the wave phase, $\chi=\omega t- k z$, where $\omega$ and $k$ are, respectively,  the frequency and wavenumber of the wave. The dispersive characteristics of the wave is insured by  $\omega\neq k$
\cite{forsberg,brodin,mendonca,bamba,asseo,chesters,gayer,polnarev,ingram,servin,mendonca2,silker}.

Substituting the above  metric in Eq.~\eqref{plasmawaveequation}, the transverse components of the electromagnetic field $A_1(t,z)=A_+(t,z)$ and $A_2(t,z)=A_-(t,z)$ obey \cite{asenjomahajanGrav}
\begin{equation}
\frac{\partial}{\partial t}\left(f_\pm \frac{\partial A_\pm}{\partial t} \right)-\frac{\partial}{\partial z}\left(f_\pm \frac{\partial A_\pm}{\partial z} \right)+\Omega_p^2 f_\pm A_\pm=0\, ,
\label{plasmawaveequation2b}
\end{equation}
where $f_+=\sqrt{-g}g^{11}\approx 1-h$,  and  $f_-=\sqrt{-g}g^{22}\approx 1+h$. 

As is evident, we could simply investigate either of the two components. Concentrating on 
 $A_-$ equation and with a variable change (to eliminate the first derivatives)
 \begin{equation}
A_-(t,z)=\exp\left(-\frac{h(\chi)}{2} \right) B(t,z)\, ,
\end{equation}
we derive
\begin{equation}
 \frac{\partial^2 B}{\partial t^2}- \frac{\partial^2 B}{\partial z^2}-\frac{\omega^2-k^2}{2}\left[  \frac{d^2 h}{d \chi^2}+ \frac{1}{2} \left(\frac{d h}{d \chi}\right)^2\right]B+\Omega_p^2   B=0\, .
\label{plasmawaveequation2}
\end{equation}
We notice that this equation is clearly non-autonomous but its coefficients depend on $t$ and $z$ only through the wave phase $\chi$. Let us, then, introduce the orthogonal coordinate 
$\varrho=k t+\omega z$ (in $t$-$z$ plane) to find an equation in  $\chi$ and $\varrho$. Since the latter is an ignorable coordinate, we could Fourier analyze
\begin{equation}
B(\varrho,\chi)=\exp\left(- i\Theta \varrho \right)a(\chi) \, ,
\end{equation}
where $\Theta$ is an arbitrary number. We  arrive at the ordinary differential equation \begin{eqnarray}
&&	\left(\omega^2-k^2\right)\left(\frac{d^2 a}{d\chi^2}-\Theta^2 a\right)-4i \omega k\Theta\frac{d a }{d\chi}\nonumber\\
&&\qquad-\frac{\omega^2-k^2}{2}\left[  \frac{d^2 h}{d \chi^2}+ \frac{1}{2} \left(\frac{d h}{d \chi}\right)^2\right]a+\Omega_p^2   a=0\, ,
\label{plasmawaveequation3}
\end{eqnarray}
describing the structure of the EM wave in a model dispersive gravitational wave background ($\omega\neq k$). In particular  specific scenarios (with $\Theta=0$), it has been shown that in this system,  gravitational waves can excite a resonant parametric instability and amplify  electromagnetic waves \cite{asenjomahajanGrav}. 

In order to explicitly prove the mirror properties of gravitational waves, we will consider almost-luminal propagation, such that $\omega/k=1-\epsilon$, with $\epsilon\lll1$. This also implies that $\chi\rightarrow k \zeta$ and $\varrho\rightarrow k \eta$, where $\zeta=t-z$ and $\eta= t+z$ are light-cone coordinates.
By assuming further that the effective frequency of the EM wave is much smaller than that of the gravitational wave  ($|(1/a)d a/d\zeta|\ll |d h/d\zeta|$), we can neglect the second derivative term simplifying the system to [$(\omega^2-k^2) \Theta\sim \epsilon k^2\Theta  \ll \Omega_p^2$]
\begin{equation}
\frac{d a}{d\zeta}+i\frac{\rho(\zeta)}{\Theta}a=0\, ,
\label{plasmawaveequation4}
\end{equation}
with
\begin{equation}
\rho(\zeta)=\frac{\Omega_p^2}{4 k}+ \frac{\epsilon}{4k} \frac{d^2 h}{d \zeta^2}+ \frac{\epsilon}{8k} \left(\frac{d h}{d \zeta}\right)^2\, ,
\end{equation}
which is readily integrated to yield
%
\begin{equation}
a(\zeta)=a_0(\Theta) \exp\left(-\frac{i }{\Theta}\int \hat\rho\,  d\zeta \right)\, ,
\end{equation}
where $a_0(\Theta)$, independent of $\zeta$, is an arbitrary function of $\Theta$. The most general solution, however, can be constructed by superimposing Fourier components in $\eta$ weighted by $a_0(\Theta)$. The general electromagnetic wave-packet solution, thus, is 
\begin{eqnarray}
A_-(\eta,\zeta)&= &\frac{\exp(-{h(\zeta)}/{2})}{\sqrt{2\pi}}\times\nonumber\\
&& \int_{-\infty}^{\infty}a_0(\Theta)\exp\left(-i\Theta k \eta-\frac{i }{\Theta}\int \rho\,  d\zeta \right) d\Theta\, ,\nonumber\\
&&
\label{soluFirstA}
\end{eqnarray}
This last result coincides with the findings of Ref.~\cite{Esirkepov} for a general luminal plasma-vacuum interface, modeled by a profile function $\rho$. It is such function that acts like a luminal mirror to electromagnetic wave propagation. However, in our case, it is the plasma immersed in a gravitational wave background which forms the medium that reflects light.

Let us now follow the argument of Ref.~\cite{Esirkepov}, and
assume an incident simple electromagnetic plane wave, represented by $\exp(i \nu \eta)$, coming from infinity (where $\Omega_p=0$ and the variations of the gravitational wave vanish). Here $\nu$ is the frequency of the incident plane wave.  For the purpose of solution \eqref{soluFirstA} be valid, then $a_0(\Theta)=\sqrt{2\pi}\delta(\nu-\Theta k)$. Therefore, when the wave interact with an almost-luminal gravitational wave background,
the general electromagnetic wave solution \eqref{soluFirstA}
becomes  $A_-(\eta,\zeta)=  \exp\left(-{h}/{2}-i S \right)$, where its phase is given by
\begin{eqnarray}
S=\nu\eta+\frac{k}{\nu}\int \rho\,  d\zeta \, .
\label{soluFirstA2}
\end{eqnarray}
Form here, we can calculate the frequency of this electromagnetic plasma wave as $W=-\partial S/\partial t=\nu+ k \rho/\nu$, and its wave number $K=\partial S/\partial z=-\nu+ k \rho/\nu$. Thereby, the phase velocity $v_{ph}=W/K$, and group velocity $v_{gr}=\partial W/\partial K$, of the  electromagnetic plasma wave interacting with the gravitational wave background are
\begin{equation}
v_{ph}=\frac{1}{v_{gr}}=\frac{\nu^2+ k\rho}{-\nu^2+ k\rho}\, .
\end{equation}
The reflection of the electromagnetic plasma wave (compared with the incident wave from infinity) is achieved when $v_{ph}>0$ and $v_{gr}>0$ \cite{Esirkepov}. This occurs when the frequency of the incident plane wave fulfills 
\begin{equation}
\nu^2<k \rho=\frac{\Omega_p^2}{4}+ \frac{\epsilon }{4} \frac{d^2 h}{d \zeta^2}+ \frac{\epsilon }{8} \left(\frac{d h}{d \zeta}\right)^2\, .
\label{conditionreflection}
\end{equation}
Notice that, because of our assumptions,  the gravitational wave contribution can be the dominant one to the reflection phenomenum (as the plasma could be very diluted).
 Under the condition \eqref{conditionreflection}, the ratio between the frequencies of the
reflected and incident wave is
\begin{equation}
\frac{W}{\nu}=1+\frac{k \rho}{\nu^2}>2\, ,
\end{equation}
which is unbounded for $W$. Therefore the reflected electromagnetic plasma wave is always upshifted by its interaction with gravitational waves. 

Thus, any very low-frequency (very long wavelength) incident electromagnetic plane wave can be reflected by the gravitational wave background suffering an increase of its frequency in the process. This phenomena could occur  in the absence of a plasma medium, when $\Omega_p\approx 0$. In such cases is evident that the reflection is produced solely by gravitational wave background.

Let us estimate the result \eqref{conditionreflection}
in order to show its relevance. Assuming a simple gravitational wave propagating in a sinusoidal manner as $h=h_0\sin(\omega \zeta)$. In order to fulfill all the above assumptions, we choose that $h_0\sim \mathcal{O}\left(\epsilon^{1/2}\right)$, and $\omega\sim \mathcal{O}\left(\epsilon^{-3/4}\right)$. This implies that the gravitational wave background has a very high frequency. Also, this implies that $|{d^2 h}/{d \zeta^2}|\sim\omega^2 h_0\gg\left({d h}/{d \zeta}\right)^2\sim\omega^2 h_0^2$. Therefore, for a frequency of the incident wave  fulfilling $\nu^2\ll \Omega_p^2+\epsilon  h_0\omega^2$ the reflection, with a large upshifting in the frequency, is guaranteed. Notice that $\epsilon  h_0\omega^2\sim \mathcal{O}\left(1\right)$. We then deduce that any high-frequency gravitational wave background  can reflect electromagnetic waves.

The main conclusion of the above calculations is that, under appropriate conditions, an electromagnetic wave traveling in a backdrop of dispersive gravitational waves, can undergo reflection; and the reflected wave is    frequency upshifted. Naturally such a scenario can pertain in all eras of the universe and one must look for the manifestation of this rather striking phenomena. Such an upshift could interfere with, for example, with the cosmic redshift measurements.  This additional mechanism for  frequency shift could  be very significant if the EM waves were originally in the low frequency part of the spectrum. It may be legitimate to suggest that in different astrophysical/cosmological measurements, the observed frequency of a signal could be rooted in a lower frequency original wave.

\section*{Acknowledgments}

FAA thanks to FONDECYT grant No. 1230094 that partially supported this work. This research is also supported by US DOE Grants. DE- FG02-04ER54742 and DE-AC02-09CH11466.

\end{document}